\documentclass[aps,pra,preprint,groupedaddress,floatfix]{revtex4-1}

\newcommand\T{\rule{0pt}{2.6ex}}
\newcommand\B{\rule[-1.2ex]{0pt}{0pt}}
\newcommand\Pmax{$P_{\text{max}}$ }
\newcommand\Pavg{$\langle P_s\rangle$ }
\newcommand{\eps}{\varepsilon}
\usepackage{graphicx}
\usepackage{epstopdf}
\usepackage{amsmath}
\usepackage{amsfonts}
\usepackage{dcolumn}
\usepackage{bbm}
\usepackage{array}
\usepackage{subfigure}

\newcolumntype{.}{D{.}{.}{2}}

\begin{document}


\title{Quantum walk-based search and centrality}


\author{Scott D.~Berry}
\email[]{berrys01@student.uwa.edu.au}
\author{Jingbo B.~Wang}
\email[]{wang@physics.uwa.edu.au}
\affiliation{School of Physics, The University of Western Australia, 6009 Perth, Australia}


\date{\today}

\begin{abstract}
We study the discrete-time quantum walk-based search for a marked vertex on a graph. By considering various structures in which not all vertices are equivalent, we investigate the relationship between the successful search probability and the position of the marked vertex, in particular its centrality. We find that the maximum value of the search probability does not necessarily increase as the marked vertex becomes more central and we investigate an interesting relationship between the frequency of the successful search probability and the centrality of the marked vertex.
\end{abstract}


\maketitle

%
\section{\label{intro} Introduction}
Quantum walks are the quantum analogue of classical random walks. Rather than stepping with a certain probability between adjacent vertices of a graph, a quantum walker is characterised by a set of probability amplitudes associated with vertices of the graph \cite{kempe2003a}. The strikingly different behaviour of quantum walks from their classical counterparts has already been harnessed in the formulation of quantum walk-based algorithms which can outperform corresponding classical algorithms \cite{kendon2006a,santha2008a,smith2010a}. While the continued interest in quantum walks is largely due to these algorithmic applications in the context of quantum computation, the quantum walk also forms a powerful and flexible model of the evolution of a coherent (or partially decoherent) quantum system \cite{katori2005a, oka2005a,kurzynski2008a,qing-kuan2009a,broome2010a}. Since analytical techniques are currently being developed to analyse quantum walks, their application to diverse problems in physics will likely become increasingly common. Given also the usefulness of classical random walks in studying transport on complex structures, it is interesting from a physical standpoint to continue to characterise quantum walks on graphs.

Searching is one of the major problems in computer science and a large amount of research in the field of theoretical quantum computation has been in the development of general algorithms for fast searching of databases. Quantum search algorithms were first introduced by Grover to search an unsorted database \cite{grover1997a,grover1997b} and later extended to quantum walk-based search algorithms for specific database topologies in both the discrete-time \cite{shenvi2003a,ambainis2005a,reitzner2009a} and continuous-time \cite{childs2004b} cases. These studies focused on highly symmetric structures such as hypercubic lattices, complete and bipartite graphs, and found that the topology of the database was crucial in determining the efficiency of the search. An important difference between the discrete and continuous-time formulations of quantum walks is the extra ``coin'' degrees of freedom in the discrete-time case. Ambainis \emph{et al.}~\cite{ambainis2005a} demonstrated that discrete-time search can achieve full quadratic speedup relative to classical search for hypercubic lattices in $ \ge 3$ dimensions and it outperforms continuous-time search for lattices in two spatial dimensions. As shown by Childs \& Goldstone \cite{childs2004b}, continuous-time search on the hypercube only achieves the full quadratic speedup for $d$-dimensional lattices where $d > 4$.

A recent paper by Agliari \emph{et al.}~\cite{agliari2010a} considers continuous-time quantum walk-based search on fractals and thus represents the first effort to characterise the search procedure on structures that are not vertex-transitive. An interesting phenomenon which arises when studying structures where not all vertices are equivalent is that the successful search probability depends on the location of the marked vertex. In their study of quantum search on Cayley trees, T-fractals and dual Sierpi\'{n}ski gaskets, Agliari \emph{et al.}~assumed that a peripheral vertex would be more difficult to find than a more central vertex, \emph{i.e.} the maximum success probability for a central vertex would be greater than for a peripheral vertex. In this work, we analyse this idea in more detail by studying how the maximum success probability varies with the centrality of the marked vertex. We find that in some simple cases, the maximum success probability does indeed increase with increasing centrality. However we show that, in general, such a relationship does not hold.

The efficiency of quantum walk-based search relative to classical search is not only determined by the maximum success probability, but also the time taken to reach the maximum. We therefore analyse the lowest frequency of the success probability as an indicator of the time complexity of the search. Our results suggest that this frequency is correlated with the centrality for a larger class of graphs than the maximum success probability and we discuss exceptions in terms of local structure of these graphs. 

In this article, we study discrete-time quantum walk-based search on non-vertex-transitive structures, and show that the frequencies present in the success probability are determined by the global structure of the graph as well as the centrality and local structure surrounding the marked vertex. To the best of our knowledge, our derivation in the Appendix contains the first analytical solution for discrete-time quantum walk-based search on a finite line with two reflecting boundaries, which provides the characteristic frequencies of the quantum walk-based search probability.

The article is organised as follows. Sec.~\ref{QWsearch} provides an introduction to quantum walks and quantum walk-based search. Sec.~\ref{centrality} describes the measure of centrality used. In Sec.~\ref{structures} we describe the structures considered and in Sec.~\ref{CT}, \ref{RHF}, \ref{SG} we present our analytical and numerical results. Finally, Sec.~\ref{conclusion} contains discussions and conclusions. In the Appendix we provide details of our analytical calculations.

\section{\label{QWsearch} Quantum walk-based search on graphs}

Let $G(V,E)$ be an undirected graph with vertex set $V = \{v_1,v_2,v_3,\hdots \}$ and edge set $E = \{(v_i,v_j),(v_k,v_l), \hdots \}$ consisting of unordered pairs of connected vertices. If there are $d$ edges incident on a vertex $v_i$, we say that $v_i$ has degree $d$. As described in \cite{aharonov2001a,kendon2005a}, $\mathcal{H_P}$ is defined as the position Hilbert space, which is spanned by an orthonormal basis of vertex states $\{ |v_i\rangle : v_i \in V \}$. For a graph of maximum degree $d$, $\mathcal{H_C}$ is defined as the $d$-dimensional coin Hilbert space spanned by the orthonormal basis of coin states $\{ |c_i\rangle : i =1,\hdots,d\}$, representing the outgoing edges at a vertex $v_i$. The discrete-time quantum walk considered here takes place on the subnodes of the graph, which are represented by product states of the form $|v\rangle \otimes |c\rangle  = |v,c \rangle \in \mathcal{H_P} \otimes \mathcal{H_C}$. Note that if $G$ is not $d$-regular, then there are vertices of degree $d_i < d$. In this case the states $\{|v_i,c \rangle : c > d_i$\} do not physically represent subnodes of the graph $G$ and are not occupied at any stage of a quantum walk on $G$.

Let one step of the discrete-time quantum walk on the graph be the application of the unitary time-evolution operator $U = S \cdot(\mathbbm{1}\otimes C)$, where $S$ is the shift operator and $C$ is the coin operator. $S$ acts on the extended position space $\mathcal{H_P}\otimes \mathcal{H_C}$ as,
\begin{equation}
S|v_i,c_j\rangle=|v_j,c_i\rangle ,
\end{equation}
where $|v_i,c_j\rangle$ is the subnode state corresponding to the edge $(v_i,v_j)$ at the vertex $v_i$. The coin operator $C$ at a vertex $v_i$ of degree $d_i$ can be represented by a $d_i \times d_i$ matrix, which mixes the probability amplitudes of the subnode states of $v_i$. We mainly consider symmetric coin matrices, so that the ordering of the subnodes at a particular vertex is unimportant. In the Appendix, when we reduce the quantum walk on a Cayley tree to a one-dimensional walk, we need a biased coin. The labelling will then become important and will be made explicit.

We follow the procedure introduced by Shenvi \emph{et al.}~\cite{shenvi2003a} for the discrete-time quantum walk-based search for a marked item. The quantum walker initially has equal probability to be found at each vertex. \emph{i.e.} the state $|\Psi_0\rangle$ is an equal superposition of all vertex states $|v_i\rangle \in \mathcal{H_P}$. The probability amplitude at each vertex is then divided equally between all subnodes. It should be noted that for a graph which is not degree-regular this is \emph{not} the same as an equal superposition of all subnode states. Formally, the initial state is given by
\begin{equation}
|\Psi_0\rangle = \frac{1}{\sqrt{N}}\sum_{i=1}^N\sum_{j=1}^{d_i} \frac{1}{\sqrt{d_i}}|v_i,c_j\rangle.
\label{psi0}
\end{equation} 
Now consider a subset $M \subset V$ of \emph{marked vertices}. The marking is intended to represent a ``quantum oracle" and is implemented as a perturbation to the coin operator at the marked vertices. A precise description of an oracle in this context can be found in \cite{shenvi2003a}. The quantum search procedure proceeds via the repeated application of the perturbed time-evolution operator, $U' = S\cdot(\mathbbm{1}\otimes C')$ where the coin operator at vertex $v_i$ is given by
\begin{equation}
(C'_i)_{mn} = \begin{cases}
\: -\delta_{mn}+2/d_i & , \: v_i \notin M \\
\: -\delta_{mn} & ,\: v_i \in M
\end{cases}
\end{equation}
\begin{equation*}
\text{ for } m,n = 1,\hdots, d_i.
\end{equation*}
The coin operator above for $v_i \notin M$ will be referred to as the Grover coin. For all examples in this paper, the set $M$ will contain only 1 vertex. The success probability, $P_s(t)$, is defined as the probability of finding the quantum walker at the marked vertex $v_m$ at time $t$. This is given by 
\begin{equation}
P_s(t) := |\langle v_m |\Psi(t)\rangle|^2 = |\langle v_m | (U')^t |\Psi_0\rangle|^2 \text{ where } v_m \in M.
\label{success}
\end{equation}
The time averaged success probability is denoted by $\langle P_s\rangle$. As shown in Eq.~\ref{success}, $P_s(t)$ is determined by unitary time-evolution from the initial state. Reversibility of unitary processes implies that $P_s(t)$ does not converge for large $t$ but instead oscillates, which allows us to define the search frequency $\omega_s$ as the lowest frequency present in the success probability. On graphs for which quantum walks are not exactly periodic, this is computed from the power spectrum of $P_s(t)$ by selecting the lowest frequency above noise. Assuming that the sample time is sufficient that the Fourier transform has converged, we define the threshold as 10\% of the highest peak present in the power spectrum. For the simple graphs studied here the power spectra computed contain only a few frequencies and this functional definition is adequate. We expect, however, that for graphs with less symmetry the threshold may need to be modified.

Using $\omega_s$ it is possible to determine the ``period'' of $P_s(t)$, that is the approximate integer time difference between minima in $P_s(t)$. To be of any use in the search context the search probability must reach a high value during the first period. We therefore quantify the success of the search as $P_{\text{max}}:=\text{max}\{P_s(t): 1 \le t \le 2\pi/\omega_s, t \in \mathbb{Z} \}$.

The quantum search procedure resembles the wavelike propagation of a (phase inverted) perturbation over a graph. The perturbation originates from the marked vertex at each time step and results in time-dependent amplitude amplification at the marked vertex, measured by $P_s(t)$. The maxima in $P_s(t)$ occur when the probability amplitudes constructively interfere at the marked vertex, which is highly dependent on the structure of the graph. The amplitude and time dependence of $P_s(t)$ are therefore determined by both the local and global structure of the graph. In turn, these quantities may be used to provide information about this underlying structure. Indeed, using a similar procedure, Douglas \& Wang \cite{douglas2008a} gave evidence that the information contained in these amplitudes was sufficient to distinguish pairs of non-isomorphic graphs for all cases tested. It is reasonable therefore to expect that the centrality of a vertex should affect both \Pmax and $\omega_s$.

\section{\label{centrality}Centrality}
We now define the measure of centrality considered in this paper. The random walk centrality (RWC), introduced by Noh \& Reiger \cite{noh2004a} is a measure of centrality designed to represent the relative speed with which a given vertex can receive and send information over a network. $P_{ij}(t)$ is defined as the probability of a classical random walker starting at the vertex $v_i$ to be at the vertex $v_j$ after a time $t$. The random walk centrality of a vertex $v_j$ is then defined as
\begin{equation}
\text{RWC}_j:=\frac{P_j^\infty}{\tau_j},
\label{eqRWC}
\end{equation}
where $\tau_j = \sum_{t=0}^\infty \{P_{jj}(t) - P_j^\infty \}$ and $P_j^\infty$ := $\lim_{t\rightarrow\infty}P_{ij}(t)$ which is the same for all $i$. $P_{ij}(t)$ is calculated using the master equation,
\begin{equation}
\label{master1}
P_{ij}(t+1) = \sum_k \frac{A_{kj}}{d_k}P_{ik}(t).
\end{equation}
Here $A$ is the adjacency matrix of the graph and $d_k$ is the degree of vertex $v_k$. 
This definition was proposed in the context of complex networks and there is an implicit assumption that the expression for $\tau_j$ converges, \emph{i.e.}~$\lim_{t\rightarrow\infty}P_{ij}(t)$ exists. The master equation (Eq.~\ref{master1}) describes a random walker that changes its location at each time step. Therefore, for graphs which contain closed walks of only even length, only sites at even distances from the start site are occupied after even times. In this case the probability distribution does not converge, which prevents the calculation of $\tau_j$ (Eq.~\ref{eqRWC}). This difficulty is easily overcome by redefining the master equation in terms of the \emph{lazy} random walk, such that the \emph{lazy} random walker only moves at each time step with probability of 1/2. This leads to the following master equation,
\begin{equation}
P_{ij}(t+1) = \frac{1}{2}\sum_k \left(\delta_{kj} + \frac{A_{kj}}{d_k} \right)P_{ik}(t).
\end{equation}
In this case, $\lim_{t\rightarrow\infty}P_{ij}(t) = d_j/N$ on all connected graphs regardless of the initial state \cite{levin2008a}.
On the Sierpi\'{n}ski gasket, where both the lazy RWC and the normal RWC are defined, we find that they show the same qualitative behaviour. For the remainder of the paper, RWC will refer to the lazy random walk centrality. RWC is an example of a closeness centrality measure, in that larger values of RWC are associated with closeness to the centre of a network \cite{borgatti2006a}.

\section{\label{structures}Topology and connectivity of graphs considered}

\begin{figure}[tb]
   \centering
   \includegraphics{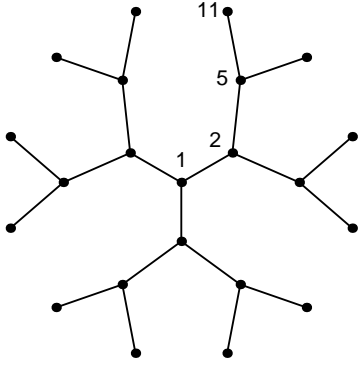}
   \caption{\label{3CT3fig}The third generation 3-Cayley tree. Vertices are ranked and labeled according to their random walk centrality. Non-labelled vertices are structurally equivalent to one of the labelled vertices.}
\end{figure}

We now describe the structures considered in this study. The $n$-th generation $d$-Cayley tree (as shown in Fig.~\ref{3CT3fig}) is a tree of $n$ levels in which all vertices on the interior have degree $d_i=d$. The outermost layer of the tree is called the \emph{surface}. All vertices on the surface are called \emph{leaves} and have $d_i = 1$. The total number of vertices in a Cayley tree is $N = (d \times (d-1)^n - 2)/ (d-2)$. The structure is defined more rigorously in the Appendix. Quantum walks on Cayley trees, or the closely-related ``glued-trees'' graph of Childs \emph{et al.}~have been extensively studied in continuous-time \cite{childs2003a,mulken2006a, agliari2008a,farhi1998a,agliari2010a}, and discrete-time \cite{chisaki2009a,tregenna2003a,carneiro2005a}, but not yet for discrete-time quantum search using marking operators.

\begin{figure}[tb]
   \centering
   \includegraphics{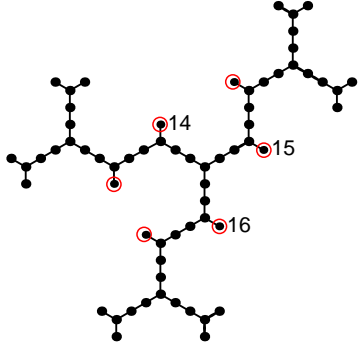}
   \caption{\label{RHFfig} (Color online) The third generation regular hyperbranched fractal of functionality, $f = 3$ (RHF$_{3,3}$). The \emph{cul-de-sac} vertices are highlighted.}
\end{figure}

A first generation regular hyperbranched fractal (RHF) (as shown in Fig. \ref{RHFfig}) of functionality $f$ is a star graph consisting of a central vertex connected through $f$ edges to $f$ surface vertices. To construct a second generation RHF, $f$ copies of the first generation RHF are connected to the core first generation RHF through a single leaf-leaf edge. This procedure is repeated $n$ times for an $n$-th generation RHF. The number of vertices in an RHF therefore grows exponentially with the generation and is given by the formula, $N = (f+1)^n$. The maximum degree of a vertex in an RHF is $f$. Star graphs (RHF$_{1,f}$) and other RHF$_{n,f}$ have been studied previously for continuous-time quantum walks in \cite{xu2009a, volta2009a}, but not for discrete-time quantum walks.

While the first generation Cayley trees and RHF$_{1,f}$ are identical, the structures are distinct for $n > 1$. An important difference in the context of this study is that for $n > 2$, RHFs contain so-called \emph{cul-de-sac} vertices, which are leaves that are connected to main paths. Note that we do not consider the leaves of the Cayley tree to be \emph{cul-de-sac} vertices since their neighbours do not lie on main paths. 

\begin{figure}[tb]
   \centering
   \includegraphics{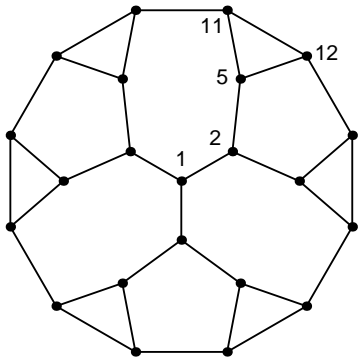}
   \caption{\label{3CT3joinpic}The third generation \emph{joined} 3-Cayley tree obtained from the equivalent Cayley tree by connecting surface vertices. Vertices are labelled as for the equivalent Cayley tree. Non-labelled vertices are structurally equivalent to one of the labelled vertices.}
\end{figure}

\begin{figure}[tb]
   \centering
   \includegraphics{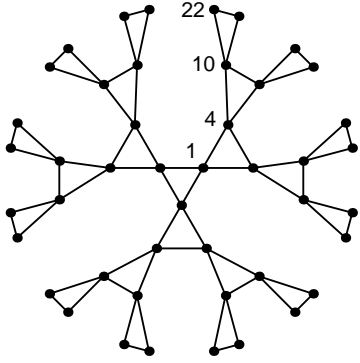}
   \caption{\label{hc43pic}The $N=45$ Husimi cactus obtained as a dual structure to the fourth generation 3-Cayley tree. Vertices are ranked and labeled according to their random walk centrality. Non-labelled vertices are structurally equivalent to one of the labelled vertices.}
\end{figure}

\begin{figure}[tb]
   \centering
   \includegraphics{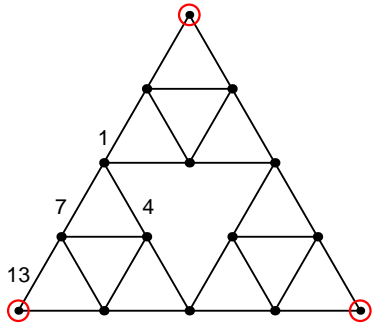}
   \caption{\label{sierp2fig}(Color online) The second generation Sierpi\'{n}ski gasket. Vertices are ranked and labeled according to their random walk centrality. Non-labelled vertices are structurally equivalent to one of the labelled vertices. Highlighted vertices are peripheral (least central as measured by RWC).}
\end{figure}

We also study structures that contain more than one simple path between all pairs of vertices. Specifically, we consider the \emph{joined} Cayley tree, the Husimi cactus and the Sierpi\'{n}ski gasket. The joined Cayley tree is obtained from the Cayley tree by adding edges between surface vertices (as shown in Fig.~\ref{3CT3joinpic}). For trees with $d=3$ this results in a $3$-regular graph. The Husimi cactus (Fig.~\ref{hc43pic}) is a dual structure to the Cayley tree, constructed by placing a vertex at each edge of the corresponding Cayley tree and connecting vertices that represent adjacent edges in the Cayley tree \cite{blumen2006a}. The second generation Sierpi\'{n}ski gasket is shown in Fig.~\ref{sierp2fig} and the structural details are explained in detail in \cite{barlow1988a}. The number of vertices of an $n$-th generation Sierpi\'{n}ski gasket is $N= \frac{3}{2}(3^n +1)$.

%
\section{\label{CT}Quantum search on Cayley trees}
Firstly, we provide an example of the success probabilities obtained for two inequivalent vertices on a Cayley tree. We then present analytical results for the success probability $P_s(t)$ for the $d$-Cayley tree when the central vertex is marked. Numerical results are then presented for $P_{\text{max}}$ and $\omega_s$ on various Cayley trees when a non-central vertex is marked. These results are compared with the centrality of the marked vertex. For all simulations in sections \ref{CT}, \ref{RHF} and \ref{SG}, the initial state is given by Eq.~\ref{psi0}.

\subsection{Quantum walk-based search characteristics}

\begin{figure*}[tb]
   \centering
   \subfigure{ \label{node1}
   \includegraphics{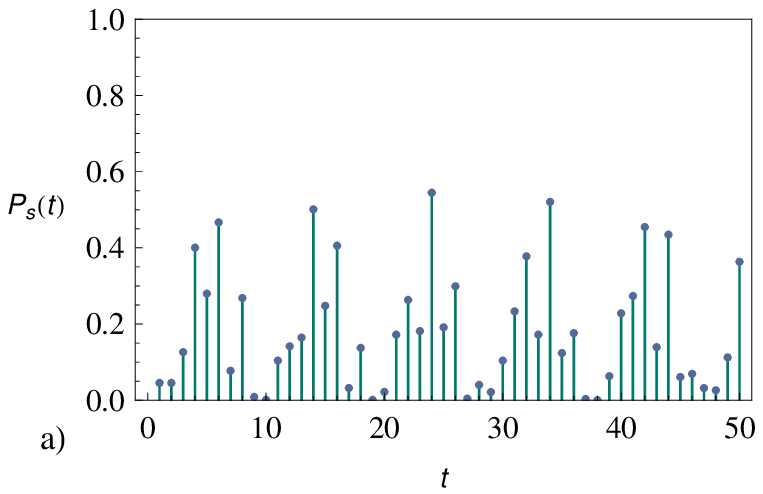}}
   \subfigure{ \label{node11}
   \includegraphics{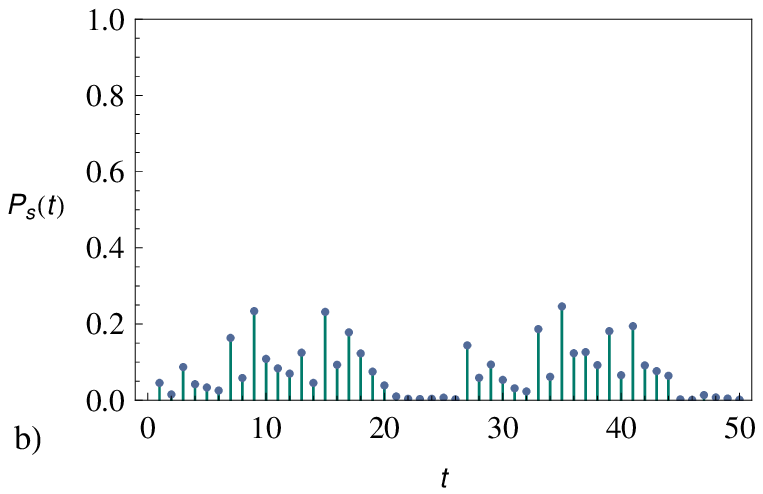}}
   \caption{(Color online) a) Numerical results for the success probability $P_s(t)$ on the third generation 3-Cayley tree (stucture shown in Fig.~\ref{3CT3fig}) for a central marked vertex (vertex 1). b) $P_s(t)$ for a peripheral marked vertex (vertex 11) on the same graph.}
   \label{3CT3prob}
\end{figure*}

We begin with an example of the success probability obtained on the third generation 3-Cayley tree (structure shown in Fig.~\ref{3CT3fig}). Fig.~\ref{node1} shows the success probability $P_s(t)$ as a function of time when the central vertex is marked. Fig.~\ref{node11} shows $P_s(t)$ for a peripheral marked vertex on the same graph. Comparing these plots, it can be seen that $P_s(t)$ is quasi-periodic in both cases, with a smaller ``period" when the central vertex is marked in comparison with a peripheral marked vertex. $P_s(t)$ also has a greater maximum and average amplitude for the central marked vertex. In the following, we examine how the lowest frequency $\omega_s$ and the maximum amplitude $P_\text{max}$ vary on Cayley trees.

\subsection{Central marked vertex}
\label{ResAnalytical}

\begin{figure*}[tb]
   \centering
   \subfigure{ \label{dCT}
   \includegraphics{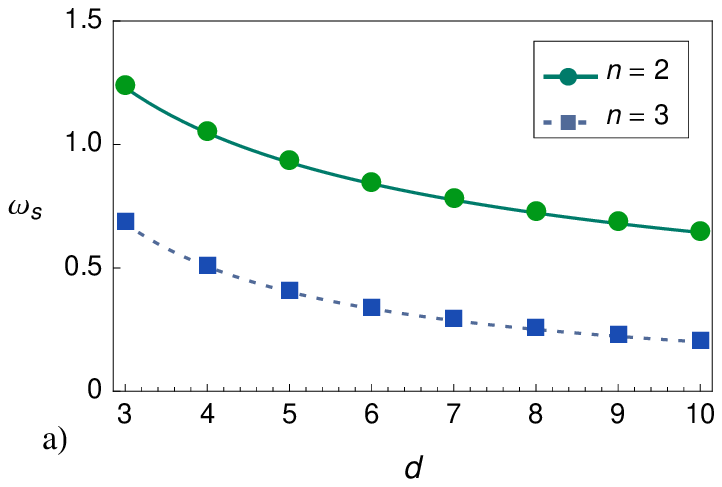}}
   \subfigure{ \label{dCTrwc}
   \includegraphics{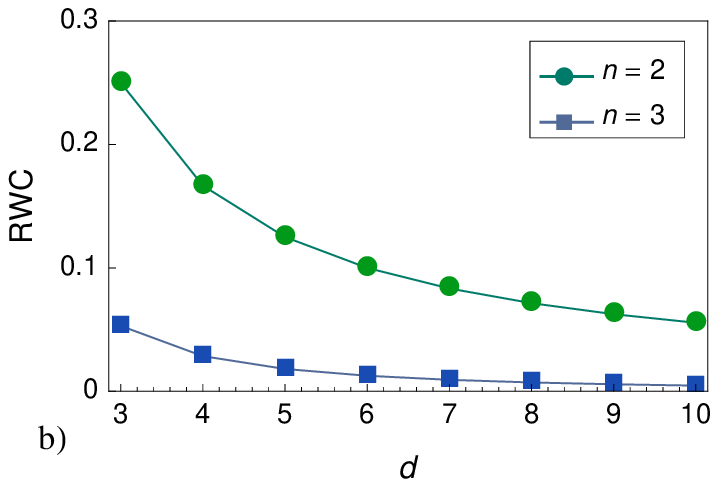}}
   \caption{(Color online) a) Analytically obtained curves for the search frequency $\omega_s(n,d)$ for $n=2$ and $3$ on a Cayley tree with a central marked vertex. Also plotted are numerical data obtained by direct simulation of discrete-time quantum search. b) Random walk centrality for the central vertex on the same Cayley trees.}
\end{figure*}

In the Appendix we show that a quantum walk on an $n$-th generation $d$-Cayley tree can be mapped to a one-dimensional quantum walk. We then study a finite one-dimensional walk with two reflecting boundaries to derive the following expression for $\omega_s$, 
\begin{align}
\label{omega_s}
\omega_s & =\omega_s(n,d):= \arctan\left(\frac{2\sqrt{d^{n-1}-1}}{d^{n-1}-2}\right) \\
& \text{ for } n = 2, 3 \text{ and } 3 \le d < \infty. \nonumber
\end{align} 

This expression is plotted in Fig.~\ref{dCT} together with our numerical results obtained by direct simulation of discrete-time quantum search. The numerical and analytical results are in perfect agreement. Finding analytical solutions for $\omega_s$ becomes more difficult for $n\ge4$ as large matrices must be diagonalised. For the second generation Cayley tree ($n=2$), we are able to derive (see Appendix) the following expression for $P_s(t)$, valid for $3\le d<\infty$,
\begin{align}
\nonumber
|\langle \mathbbm{1}|\Psi(t)\rangle|^2 = &\frac{1}{4(1+d^2)}\big\{1+d^2 + (d-1)^2\cos(\pi t) \\ 
\nonumber
& - (d^2-1)\cos(\omega_s t)+ 2\sqrt{d(d-1)}\sin(\omega_s t) \\
\nonumber
& - (d^2-2d-1)\cos[(\pi-\omega_s)t] \\ 
& - 2\sqrt{d(d-1)}\sin[(\pi-\omega_s)t] \big\}.
\label{PsCT2}
\end{align}
As can be seen from Eq.~\ref{PsCT2}, $P_s(t)$ contains only three frequencies, $\pi$, $\omega_s$ and $\pi - \omega_s$. It is interesting to note that $\omega_s(2,4) = \pi/3$, which means the only frequencies present in $P_s(t)$ are $\{\pi/3, 2\pi/3, \pi\}$. Thus $P_s(t)$ is exactly periodic with period 6. 

The search frequencies $\omega_s(2,d)$ and $\omega_s(3,d)$ are plotted in Fig.~\ref{dCT}. Although for fixed $n$ the distance between the central node and the surface is fixed, it can be seen from Fig.~\ref{dCT} that $\omega_s(n,d)$ decreases monotonically with increasing branching rate. According to Eq.~\ref{omega_s}, for large $d$, $\omega_s$ tends toward $0$. This is consistent with the results of Carneiro \emph{et al.}~\cite{carneiro2005a}, who generalised a discrete-time formulation of Childs' \emph{glued-trees} algorithm \cite{childs2003a} to include arbitrary branching rate $d$ and found that as $d \rightarrow \infty$, a quantum walker initially localised at the central node of the tree oscillates between the central node and the first level of the tree. While it is not strictly valid to compare absolute values of RWC between different graphs, we find numerically that RWC of the central vertex of a Cayley tree is also a monotonically decreasing function of $d$ for fixed $n$, as shown in Fig.~\ref{dCTrwc}. In this sense, $\omega_s$ shows the same behaviour as RWC for the central marked vertex of a Cayley tree.

\begin{figure*}[tb]
   \centering
   \subfigure{\label{nCT}
   \includegraphics{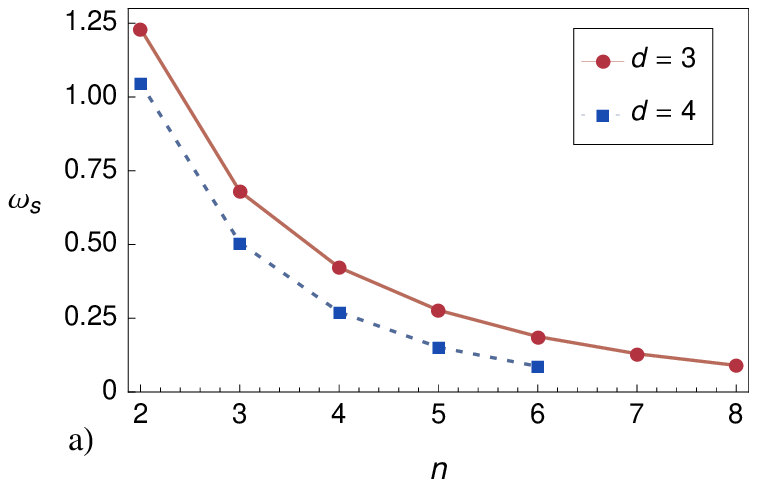}}
   \subfigure{ \label{nCTrwc}
   \includegraphics{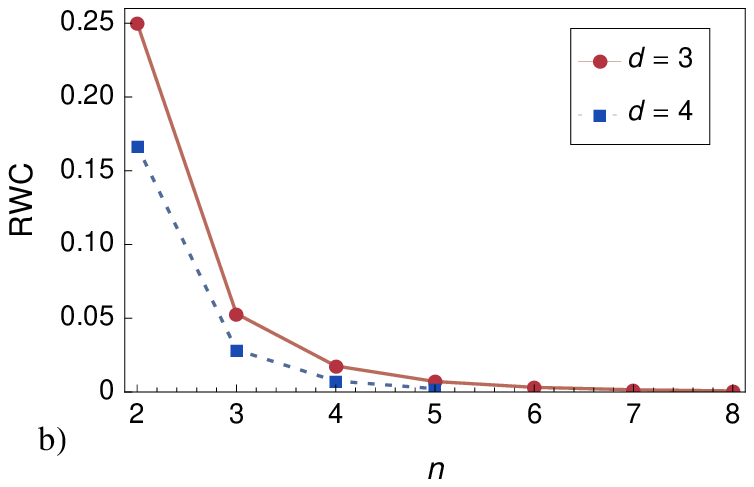}}
   \caption{(Color online) a) Numerical results for the search frequency $\omega_s$ on a Cayley tree with central marked vertex. $\omega_s$ is plotted as a function of the number of generations $n$, for $d=3$ (solid/circles) and $d=4$ (dashed/squares). b) Random walk centrality for the central vertex on the same Cayley trees.}
\end{figure*}

For $n \ge 4$ we only have numerical results for $\omega_s$, which are shown in Fig.~\ref{nCT}. The search frequency $\omega_s$ on a Cayley tree with a central marked vertex decreases as the number of generations in the tree increases. Thinking of the search procedure as generating a phase-inverted perturbation at the centre of the graph at each time step, these perturbations must constructively interfere at the centre of the graph in order to produce a maximum in $P_s(t)$. To constructively interfere, these perturbations must be reflected at the surface and thus the time between maxima in $P_s(t)$ depends on the distance between the central vertex and the surface of the graph, which increases with the number of generations, $n$. As shown in Fig.~\ref{nCTrwc}, RWC of the central vertex decreases monotonically with $n$ for fixed $d$ and thus displays the same behaviour as $\omega_s$ for the Cayley tree. 

\begin{figure}[tb]
   \centering
   \includegraphics{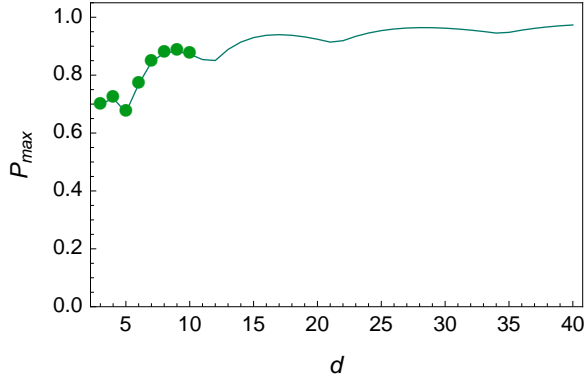}
   \caption{(Color online) Analytical results for $P_\text{max}$ on a second generation Cayley tree with a central marked node as a function of branching rate $d$ (solid line). Also plotted are numerical results (circles) obtained by direct simulation of quantum search for $3\le d \le 10$.}
   \label{PmaxCT}
\end{figure}

We now study how $P_\text{max}$ varies with $d$ for a central marked vertex on the second generation Cayley tree. By evaluating Eq.~\ref{PsCT2} at integers $t \approx \pi / \omega_s(2,d)$, we can find $P_\text{max}$ for arbitrary $d$. The results are shown in Fig.~\ref{PmaxCT} for $3\le d \le 40$. Numerical results agree with our analytical results for $3 \le d \le 10$. Fig.~\ref{PmaxCT} shows that $P_\text{max}$ has a much more complex dependence on $d$ than was seen for $\omega_s$. This complex dependence on $d$ arises because the frequencies $\pi$, $\omega_s(d)$ and $\pi - \omega_s(d)$ present in $P_s(t)$ do not necessarily constructively interfere within the first period for all $d$. While RWC and $\omega_s$ both decrease with increasing branching rate, we see that $P_\text{max}$ generally becomes larger as $d$ increases. This can be understood as arising from the greater localisation of the quantum walk close the central vertex as branching rate increases. 

\subsection{Non-central marked vertices}
\label{nonCentral}

The analysis in the Appendix is dependent on being able to map the quantum walk on a Cayley tree to a walk on a line. When a non-central vertex is marked, the same mapping cannot be used and we do not have analytical results.

\begin{figure}[tb]
   \centering
   \includegraphics{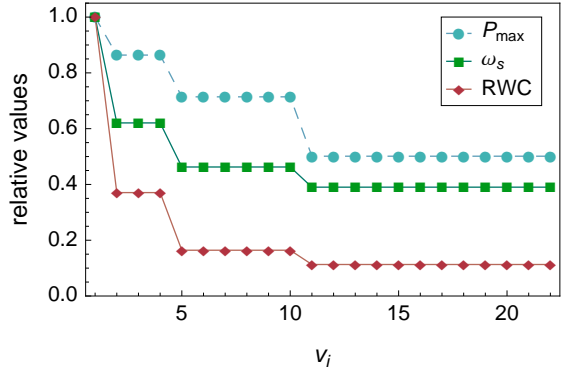} 
   \caption{(Color online) Numerical results for $P_\text{max}$, $\omega_s$ and RWC on the third generation 3-Cayley tree (all data normalised for comparison). Data plotted as a function of vertex number, $v_i$.}
   \label{3CT3res}
\end{figure}

Instead we present numerical results for $P_{\text{max}}$ and $\omega_s$ on Cayley trees where the marked vertex is not necessarily central. $P_s(t)$ was computed via direct application of $U$ to the probability amplitudes of the subnode states $|v_i,c_j\rangle $ on which the quantum walk takes place. Numerical methods were then used to compute the discrete Fourier transform and $\omega_s$ was obtained as defined in Sec~\ref{QWsearch}. For a given graph, this was repeated for all possible positions of the marked vertex. Fig.~\ref{3CT3res} shows the dependence of $P_\text{max}$ and $\omega_s$ on the centrality of the marked vertex for the third generation Cayley tree with $d=3$ (structure shown in Fig.~\ref{3CT3fig}). As can be seen in Fig.~\ref{3CT3res}, a more central vertex on Cayley tree gives rise to a success probability $P_s(t)$, which has a larger minimum frequency, $\omega_s$, and attains a greater maximum amplitude, $P_{\text{max}}$. The results for all other Cayley trees studied were analogous. 

On Cayley trees, RWC becomes smaller for vertices which are further from the central vertex because a random walker starting from a less central vertex takes, on average, longer to visit all vertices. We studied numerically the following Cayley trees: $\{d =3, n \in [2,8] \}$, $\{d \in [4,5], n \in [2,5] \}$ and $\{d \in [6,10], n \in [2,3]\}$. For all Cayley trees tested we found that $P_\text{max}$, \Pavg and $\omega_s$ decrease monotonically with RWC.

These results are consistent with those observed by M\"{u}lken \emph{et al.}~\cite{mulken2006a} who modelled exciton transport on the Cayley tree by continuous-time quantum walk and found that excitations which were initially centrally located propagated throughout the structure much more rapidly than excitations which were initially peripheral.

\section{\label{RHF}Quantum search on regular hyperbranched fractals}

\begin{figure}[tb]
   \centering
   \includegraphics{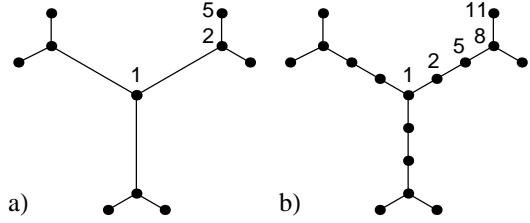}
   \caption{ \label{RHF_CT} A comparison of the structures of a) the second generation 3-Cayley tree and b) RHF$_{2,3}$. Vertices are ranked and labelled in order of decreasing random walk centrality.}
\end{figure}

\begin{figure}[tb]
   \centering
   \includegraphics{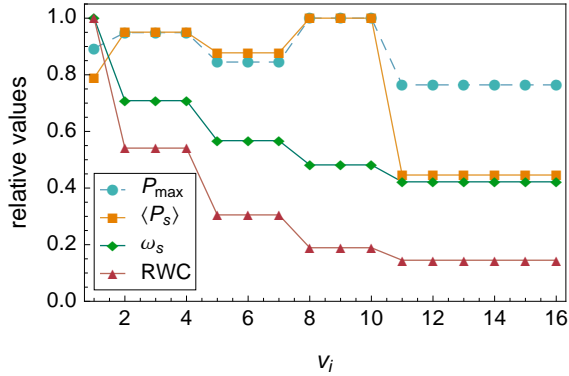} 
   \caption{\label{f3g2resa} (Color online) Numerical results for $P_\text{max}$, $\omega_s$, and $\langle P_s \rangle$ on RHF$_{2,3}$. All data is normalised for comparison and vertices $v_i$ are ordered with decreasing random walk centrality.} 
\end{figure}

We now study RHF in order to investigate how greater structural complexity affect both $P_\text{max}$  and $\omega_s$.

Firstly, we consider the simple case of $n=2, f=3$ (RHF$_{2,3}$). As shown in Fig.~\ref{RHF_CT}, the only difference between this fractal and the 2nd generation 3-Cayley tree is extra vertices of degree 2 on each branch, between the central vertex and level $1$.  In this case, like on Cayley trees, we find that $\omega_s$ decreases monotonically with RWC. However, as shown in Fig.~\ref{f3g2resa}, neither $P_\text{max}$ nor \Pavg decrease with centrality. In fact there are non-central vertices in RHF$_{2,3}$ at which $P_\text{max}$ and \Pavg attain a greater value than at the central vertex. This implies that $P_\text{max}$ and \Pavg are not determined by the centrality of the vertex. It is remarkable to consider that the addition of these 6 vertices symmetrically about the central vertex can perturb the quantum walk dynamics to such an extent that the the relationship between centrality and $P_\text{max}$ seen on Cayley trees is no longer maintained.

\begin{figure*}[tb]
   \centering
   \includegraphics{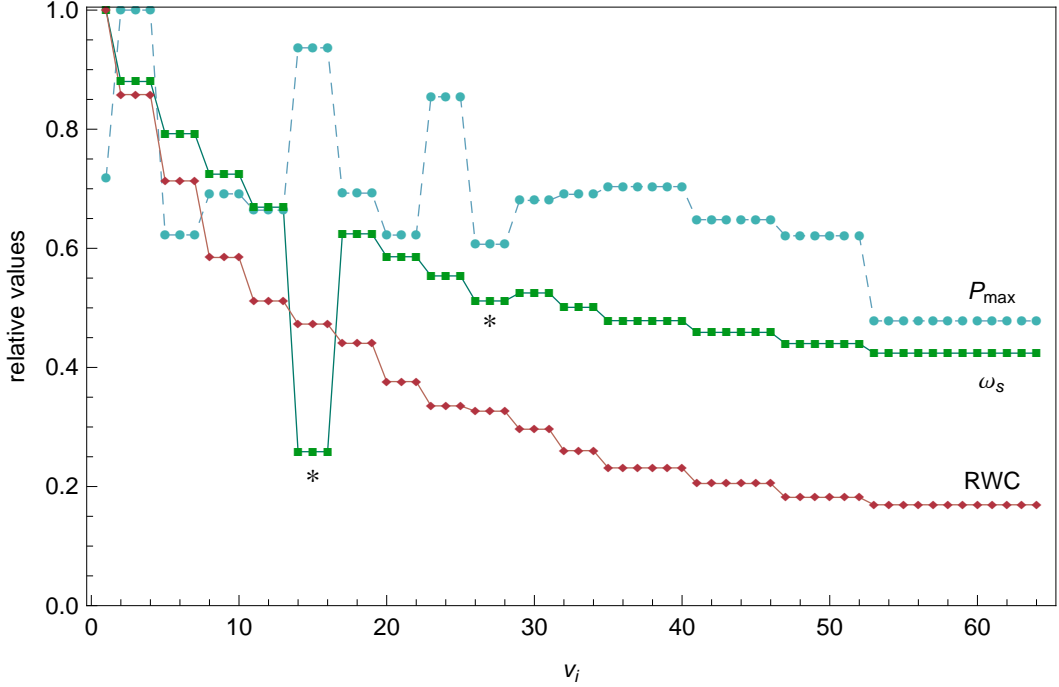} 
   \caption{\label{f3g3res}(Color online) Numerical results for $P_\text{max}$, $\omega_s$ and RWC on RHF$_{3,3}$. All data is normalised for comparison and vertices $v_i$ are ordered with decreasing random walk centrality. The ($\ast$) denotes sets of \emph{cul-de-sac} vertices.}
\end{figure*}

Fig.~\ref{f3g3res} shows the results of the calculations for $n=3,f=3$ (RHF$_{3,3}$ - structure shown in Fig.~\ref{RHFfig}). These results show that the overall trend is again decreasing $\omega_s$ with RWC but that the decrease is not monotonic. This implies that the search frequency on RHF is not solely determined by the centrality for some vertices. The vertices which break the trend have a particular structure, they are vertices which are connected by a single edge to a vertex which lies on a main path of the graph, so-called \emph{cul-de-sac} vertices (highlighted in Fig.~\ref{RHFfig}). The much smaller than expected values for $\omega_s$ at these \emph{cul-de-sac} vertices is the first evidence that the local structure surrounding the marked vertex can greatly affect the frequency of the success probability. We now investigate this further.

We construct the graph RHF$^+_{3,3}$ by adding a single extra vertex to RHF$_{3,3}$ attached through a single edge to vertex 16, which is one of the \emph{cul-de-sac} vertices in RHF$_{3,3}$ (see Fig.~\ref{RHFfig}). The calculated results of $P_\text{max}$, $\omega_s$ and RWC for RHF$_{3,3}$ and RHF$^+_{3,3}$ are shown in Table \ref{tablef3g3}. Note that in RHF$_{3,3}$, vertices $14,15$ and $16$ are structurally equivalent and have the same values for  $P_\text{max}$, $\omega_s$ and RWC. The additional vertex attached to vertex 16 in RHF$^+_{3,3}$ results in decreasing RWC for that vertex. Given that RWC is calculated using the relaxation time, this is expected. Since centrality is a relative measure among vertices, it is more informative to compare values within the same graph. We therefore compare values for vertex 16 with those for vertices 14 and 15, which are all equivalent in RHF$_{3,3}$. For RHF$^+_{3,3}$, we compute the ratios,
\begin{eqnarray*}
\frac{P_\text{max}^{(16)}}{P_\text{max}^{(14,15)}}= & \:1.02, \\ 
\frac{\omega_{s}^{(16)}}{\omega_{s}^{(14,15)}}= & \:0.99, \\ 
\frac{\text{RWC}^{(16)}}{\text{RWC}^{(14,15)}} = & \:0.96, 
\end{eqnarray*}
and we see that in real terms, the effect of the node addition to vertex $16$ is to decrease both $\omega_s$ and RWC. So we see again that $\omega_s$ appears to behave like RWC.

\begin{table}[tb]
 \begin{ruledtabular}
 \begin{tabular}{l c c r}
 & RHF$_{3,3}$ & RHF$^+_{3,3}$ & change \\ [0.5ex]
 \hline
  \multicolumn{4}{l}{$v_i = 14,15$\T\B} \\
 \hline
 $P_\text{max}$ & $2.065 \times 10^{-1}$ & $2.023 \times 10^{-1}$ & $ -2.0 \%$\\
 $\omega_s$ & $4.533 \times 10^{-2}$ & $4.574 \times 10^{-2}$ & $+0.9 \%$ \\
 RWC &  $1.166 \times 10^{-3}$ & $1.147 \times 10^{-3}$ & $-1.7 \%$ \\
 \hline
 \multicolumn{4}{l}{$v_i = 16$\T\B} 
 \\\hline
  $P_\text{max}$ & $2.065 \times 10^{-1}$ & $2.077 \times 10^{-1}$ & $ +0.6 \%$\\
 $\omega_s$ & $4.533 \times 10^{-2}$ & $4.533 \times 10^{-2}$ & 0.0\% \\
 RWC & $1.166 \times 10^{-3}$ &  $1.106\times 10^{-3}$ & $-5.4 \%$ \\
  \end{tabular}
 \end{ruledtabular}
 \caption{ \label{tablef3g3}Changes in the values of $P_\text{max}$, $\omega_s$ and RWC between RHF$_{3,3}$ and RHF$^+_{3,3}$, which has an extra vertex added to vertex 16 (see Fig.~\ref{RHFfig} for structure of RHF$_{3,3}$).}	
 \end{table}

These data suggest that, on symmetric trees, $\omega_s$ is largely determined by RWC but the local structure surrounding the marked vertex \emph{can} have a dramatic effect. It should be noted that all vertices on RHF which are observed to break the trend of decreasing $\omega_s$ with centrality have degree 1.

\section{\label{SG}Quantum search on structures with simple cycles}

\begin{figure}[tb]
   \centering
   \includegraphics{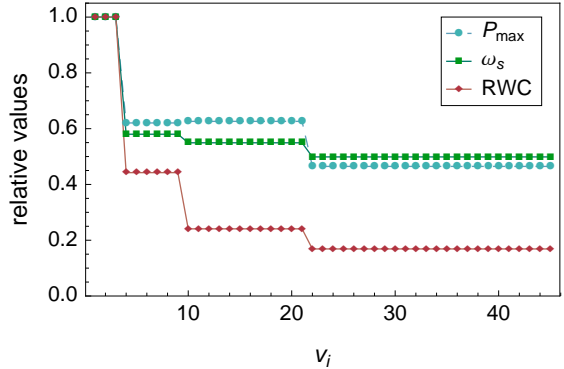} 
   \caption{\label{hc43res}(Color online) Numerical results for $P_\text{max}$, $\omega_s$ and RWC on the Husimi cactus on $N=45$ vertices. All data is normalised for comparison and vertices $v_i$ are ordered with decreasing RWC.}
\end{figure}

We now move away from trees to consider quantum walk-based search on structures which contain simple cycles. In particular, we consider the Husimi cactus, the \emph{joined} Cayley tree and a Sierpi\'{n}ski gasket. 

While a tree contains no simple cycles, a cactus is a graph in which any edge belongs to at most one simple cycle. In this sense, cacti are the most ``tree-like" graphs which contain simple cycles. On Husimi cacti derived from 3-Cayley trees, all simple cycles have length~$= 3$. Studies of continuous-time quantum walks on these Husimi cacti have found that they display similar dynamics to Cayley trees \cite{blumen2006a}. Fig.~\ref{hc43res} shows $\omega_s$, $P_\text{max}$ and RWC for the $N=45$ Husimi cactus (structure shown in Fig.~\ref{hc43pic}). Like on the Cayley tree, it can be seen that the search frequency $\omega_s$ decreases with the centrality of the marked vertex. However, in contrast to the Cayley tree, $P_\text{max}$ does not monotonically decrease with RWC. The results for the $N=21$ Husimi cactus are analogous.

\begin{figure}[tb]
   \centering
   \includegraphics{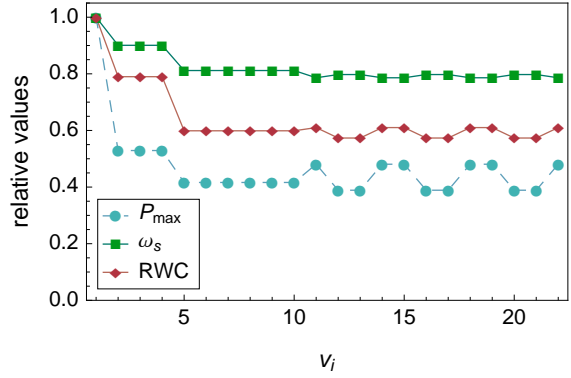} 
   \caption{\label{3CT3joinres}(Color online) Numerical results for $P_\text{max}$, $\omega_s$ and RWC on the third generation joined 3-Cayley tree (all data normalised for comparison). Data plotted as a function of vertex number, $v_i$, which is derived from the numbering of the unjoined Cayley tree. Note that vertices 11 and 12 are no longer equivalent in the joined Cayley tree.}
\end{figure}

Fig.~\ref{3CT3joinres} shows $\omega_s$, $P_\text{max}$ and RWC for the joined Cayley tree of generation 3, with $d=3$. We again see that $\omega_s$ and $P_\text{max}$ generally decrease with centrality in a similar manner to the unjoined Cayley tree. In the joined graph (Fig.~\ref{3CT3joinpic}), vertices 11 and 12 are no longer structurally equivalent since vertex 11 lies in simple cycles of length~$=\{ 3,7\}$ and vertex 12 lies in simple cycles of length~$=\{ 3,5\}$. It can be seen from Fig.~\ref{3CT3joinres} that the search procedure for marked vertices 11 and 12 now produces different values for RWC, $P_\text{max}$ and $\omega_s$. It should be noted that, for these vertices, $P_\text{max}$ and RWC display the opposite trend to $\omega_s$. This is different to the correlation observed on trees and Husimi cacti, where we saw that $\omega_s$ was usually highly correlated with RWC.

\begin{figure}[tb]
   \centering
   \includegraphics{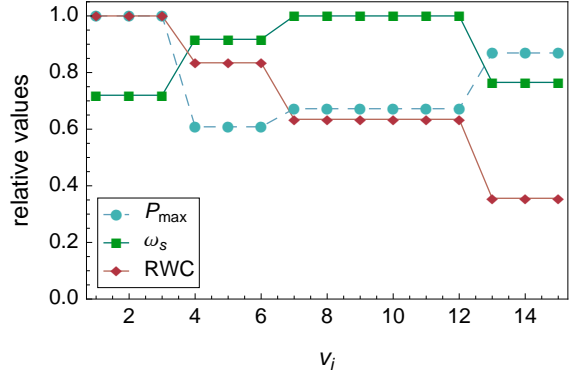} 
   \caption{\label{sierp2res}(Color online) Numerical results for $P_\text{max}$, $\omega_s$ and RWC on the second generation Sierpi\'{n}ski gasket. All data is normalised for comparison and vertices $v_i$ are ordered with decreasing RWC.}
\end{figure}

The results in Fig.~\ref{sierp2res} for the Sierpi\'{n}ski gasket show a much more complex relationship between $P_\text{max}$, $\omega_s$ and centrality to that seen on the other structures studied. We expect that the presence of simple cycles of various lengths in the structure results in interesting interference effects for quantum walks on these graphs. This interference produces much more complex behaviour of $P_s(t)$, which makes any relationship between $P_s(t)$ and the centrality of the marked vertex harder to observe. It should be noted from Fig.~\ref{sierp2res} that the peripheral (least central) vertices do not produce the smallest values of $P_\text{max}$ or $\omega_s$ in this case. 

\section{\label{conclusion}Discussion \& conclusions}

We have demonstrated that the quantum search procedure for a marked vertex on a Cayley tree results in a time-dependent success probability $P_s(t)$ that attains a greater maximum value and has a higher minimum frequency for marked vertices which are more central. Our study was extended to regular hyperbranched fractals where it was found that the maximum value of $P_s(t)$ does not necessarily decrease with centrality. We also found that the minimum frequency of the search probability $\omega_s$ on RHF is strongly related to centrality and that exceptions to this trend are caused by the \emph{local} structure surrounding the marked vertex. We therefore conclude that the success probability $P_s(t)$ for a marked vertex on a highly symmetric tree contains information about the global structure of the tree and the overall position of the marked vertex within the tree. It also contains information regarding the local structure surrounding the marked vertex.

We say that two vertices are \emph{structurally equivalent} if they have the same structural relationship to all of the other vertices of the graph (\emph{e.g.}~on a Cayley tree, all vertices which are a given distance from the centre are equivalent). We then note that all equivalent vertices on the graphs studied in this work have the same values of $P_\text{max}$ and $\omega_s$ and all inequivalent vertices have different values, suggesting that $P_\text{max}$ and $\omega_s$ could be used to partition the vertex set into structural equivalence classes. It should be noted however that this partitioning would not be possible for all graphs, since strongly regular graphs contain inequivalent vertices that produce identical success probabilities. These graphs therefore give identical values for $\omega_s$ and $P_\text{max}$ for inequivalent vertices. We find that for all vertices of degree $> 1$ on the trees and cacti studied here, $\omega_s$ can be used to order the vertices with decreasing centrality, as measured by RWC. 

The random walk centrality considered in this paper is an example of a \emph{closeness} centrality measure, \emph{i.e.} it measures the closeness of a vertex to the centre of the graph. There are, however, other classes of centrality measures \cite{borgatti2006a}. One of these is \emph{betweeness} centrality, which represents the capacity of a vertex in a network to withhold information if it were removed from the network \cite{borgatti2006a}. While betweeness and closeness are equivalent on a Cayley tree, this is not true for RHF. As seen in the studies on RHF, the values of $\omega_s$ obtained for the \emph{cul-de-sac} vertices were unusually low when compared with the corresponding values of RWC. This could be related to the low betweeness centrality of these vertices. It would be an interesting subject for further study to investigate  the relationship between characteristics of $P_s(t)$ and other measures of centrality.

It is also worth noting that $P_s(t)$ can be classically computed with time complexity $O(t N^2 \log N)$  where $t$ is the simulation time and $N$ is the number of vertices. Therefore, $\omega_s$ can be classically computed with time complexity $O(t^2\log t \,N^2\log N)$. Numerically, we find that for a $1\%$ error in $\omega_s$ for typical values of $\omega_s$ on a 3-Cayley tree, $t \approx O(N^{0.7})$ and thus the time complexity becomes $O(N^{3.4}(\log N)^2)$. This is comparable with the complexity of the equivalent RWC calculation which scales as $O(tN^2) \approx O(N^{3.1})$.

It is evident that a tremendous amount of information about the structure of the graph, and the overall position of the marked vertex is present in $P_s(t)$ and we suggest that the ``lowest frequency'' considered here is perhaps the simplest example of extracting this information. We propose that it would be extremely interesting to consider more detailed Fourier analysis of $P_s(t)$ to uncover the origin of the other frequencies present and hence determine what other information is obtainable. It would also be interesting to determine if the centrality information, which was easy to obtain via $\omega_s$ in the case of simple symmetric trees, can be extracted for more complex cases.

\begin{acknowledgments}
The authors would like to thank Brendan Douglas for helpful discussions as well as the referee for constructive comments, in particular the suggestion to study the Husimi cactus.  SDB would also like to acknowledge the financial support of John and Patricia Farrant.
\end{acknowledgments}

\appendix*
\section{Analytical results for $\omega_s$ on Cayley trees.}
\label{CTcalc}

\begin{figure*}[tb]
\includegraphics[scale=0.75]{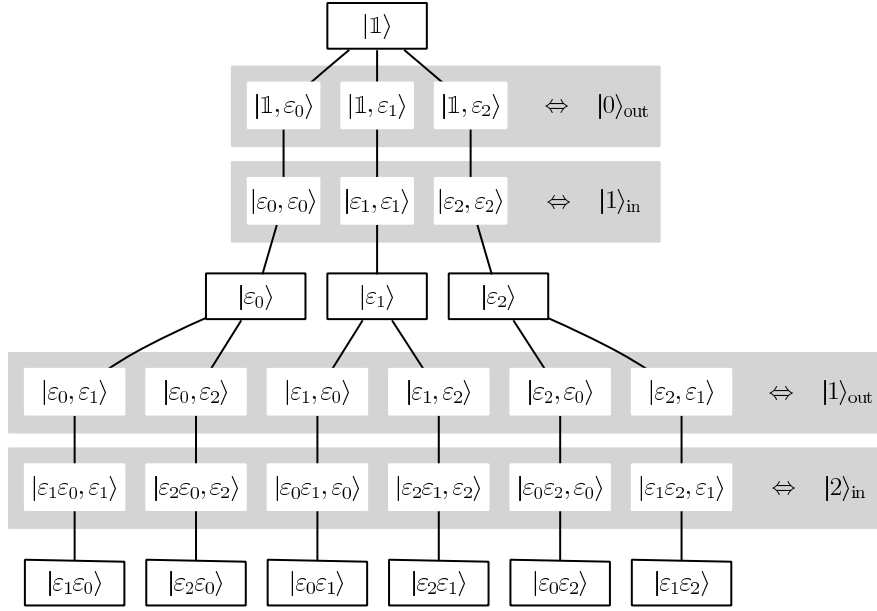}
\caption{The second generation 3-Cayley Tree, shown with the vertex states $|g\rangle \in \mathcal{H_P}$ and the subnode states $|g,\eps\rangle \in \mathcal{H_P}\otimes \mathcal{H_C}$. The shaded regions enclose groups of subnode states that make up $\{ |x\rangle_{\text{out}}, |x\rangle_{\text{in}}\}$, shown on the right.}
\label{CTfig}
\end{figure*}

We map the quantum walk on a CT to a finite one-dimensional quantum walk and derive an expression for $\omega_s$ when the central vertex is marked for $n=2, 3$. The first part of the analysis follows the method of Chisaki \emph{et~al.}~\cite{chisaki2009a} however the time-evolution operator used here is modified to incorporate the marking operator and reflection at the surface vertices (Chisaki \emph{et al.}~consider an infinite tree). The initial state is also different, therefore the complete derivation is presented.

\subsection{Defining the discrete-time quantum walk on a Cayley tree} 
An $n$-th generation Cayley tree of order $d$  is an undirected graph in which every vertex is connected to $d$ others except for vertices at a distance of $n$ from the centre. The vertex set, $V$ is defined as the group of elements generated by the set $\Sigma = \{\eps_0,\hdots,\eps_{d-1}\}$ with the constraint ${\eps_i}^2=\mathbbm{1}$, $i=0,\hdots,d-1$, 
\begin{eqnarray*}
V= & \: \{ \eps_{i_k}  \eps_{i_{k-1}} \hdots \eps_{i_2}  \eps_{i_1} \: : \: 0 \le k \le n ,\: \eps_{i_j} \in \Sigma, \\
& \: \text{ and } i_{j+1} \ne i_j  \: \text{ for } \: j=1,2,\hdots,k-1\}.
\end{eqnarray*}
The \textit{reduced word length}, $|g|$ of a vertex $g$ is the number of elements used when $g$ is written as a reduced product of $\eps_i \in \Sigma$ (e.g. $|\eps_2 \eps_2 \eps_1 \eps_0 \eps_2| =|\mathbbm{1} \eps_1 \eps_0 \eps_2| = |\eps_1 \eps_0 \eps_2| = 3$). Using this construction, the vertices at level $k$ in the tree have $|g| = k$. Vertices $g$ and $h$ are connected if and only if $gh^{-1} \in \Sigma$. This is illustrated in Fig.~\ref{CTfig}.

The quantum walk takes place on the subnodes of the graph which belong to the Hilbert space $\mathcal{H_P}\otimes\mathcal{H_C}$. We write these states $\{| g , \eps_j\rangle :  g \in V, \eps_j \in \Sigma\}$. In this basis the action of the unitary Grover coin and shift operators which drive the quantum walk can be written as,
\begin{align}
(\mathbbm{1} \otimes C) |g,\eps\rangle =  & \begin{cases}
\: \sum_{\tau \in \Sigma} (-\delta_{\eps \tau}+2/d)|g,\tau\rangle & , \: |g| < n\\
\: |g,\eps \rangle &, \: |g| = n 
\end{cases}\label{coin}\\
S |g,\eps\rangle =  & \: | \eps g,\eps\rangle.
\label{shift}
\end{align}
The coin operator (Eq.~\ref{coin}) applies a Grover coin to all vertices except those at the surface which are left unchanged, while the shift operator (Eq.~\ref{shift}) swaps the probability amplitudes between connected subnodes. With the perturbed time evolution operator $U':=S (\mathbbm{1} \otimes C')$, one step of the discrete-time quantum walk on the Cayley tree becomes,
\begin{equation}
U' |g,\eps\rangle = \begin{cases}
\: \sum_{\tau \in \Sigma} (-\delta_{\eps \tau}+2/d)|\tau g,\tau\rangle & , \: |g| < n, g \notin M \\
\: |\eps g,\eps \rangle &, \: |g| = n, g \notin M \\
\: -|\eps g,\eps \rangle &, \: g \in M .
\end{cases}
\end{equation}

\subsection{Mapping the quantum walk on a Cayley tree to a quantum walk on a finite line} 

As described in \cite{carneiro2005a,childs2003a,chisaki2009a,mulken2006a}, in certain cases, the symmetry of the initial state allows us to map the quantum walk on a Cayley tree to a walk on a line. Define the sets
\begin{align*}
E^+(x) = & \: \{ (g,\eps) \in V \times \Sigma : |g|=x,|\eps g|=x+1\}, \\
E^-(x) = & \: \{ (g,\eps) \in V \times \Sigma : |g|=x,|\eps g|=x-1\},
\end{align*}
and consider the subspace $\mathcal{H'} \subset \mathcal{H}$ spanned by following states,
\begin{align*}
|x\rangle_{\text{out}} = & \: \frac{1}{\sqrt{d(d-1)^x}}\sum_{(g,\eps) \in E^+(x)} |g,\eps\rangle, & 0 \le x \le n-1\\
|x\rangle_{\text{in}} = & \: \frac{1}{\sqrt{d(d-1)^{x-1}}}\sum_{(g,\eps) \in E^-(x)} |g,\eps\rangle, & 1 \le x \le n .
\end{align*}
as shown in Fig.~\ref{CTfig}.
If the initial state $|\Psi_0\rangle$ of the quantum walk on a Cayley tree is an equal superposition of all vertex states, equally divided among the subnodes of each vertex,
\begin{equation}
|\Psi_0\rangle  = \sqrt{\frac{d-2}{(d \times (d-1)^n - 2)}}\sum_{g \in V} \sum_{\eps :  |\eps g| < n} \frac{1}{\sqrt{d(g)}} |g, \eps \rangle,
\end{equation}
where $d(g)$ is the degree of the vertex, then $| \Psi_0\rangle$ can be written as a superposition of $\{ |x\rangle_{\text{out}}, |x\rangle_{\text{in}}\}$. 
\begin{align}
\label{initial}
|\Psi_0\rangle &= \sqrt{\frac{d-2}{d \times (d-1)^n - 2}} \bigg(\, \sum_{x=0}^{n-1}  \sqrt{(d-1)^x}|x \rangle_{\text{out}} \\
&+ \sum_{x=1}^{n-1} \sqrt{(d-1)^{x-1}}|x \rangle_{\text{in}} + \sqrt{d(d-1)^{n-1}}|n \rangle_{\text{in}} \bigg) .\nonumber
\end{align}

We now specialise to the case where there is a single marked vertex at the centre of the tree ($M = \{\mathbbm{1}\}$) and consider the action of $U'$ on the subspace $\mathcal{H'} \subset \mathcal{H_P}\otimes\mathcal{H_C}$.

For $1 \le x \le n$,
\begin{align}
\label{UprimeIn}
U' |x\rangle_{\text{in}} = & \: (\frac{2}{d}-1)|x-1\rangle_{\text{out}}+ \frac{2\sqrt{d-1}}{d} |x+1\rangle_{\text{in}},\\
\label{UprimeOut}
U' |x\rangle_{\text{out}} = & \: -(\frac{2}{d}-1)|x+1\rangle_{\text{in}}+ \frac{2\sqrt{d-1}}{d} |x-1\rangle_{\text{out}},
\end{align}
and on the boundaries,
\begin{align}
\label{Uprime0}
U' |0\rangle_{\text{out}} = & \: -|1\rangle_{\text{in}},\\
\label{Uprime1}
U'|n\rangle_{\text{in}} = & \: |n-1\rangle_{\text{out}}.
\end{align}
Note that $|\Psi_0\rangle \in \mathcal{H'} \Rightarrow |\Psi(t)\rangle = (U')^t |\Psi_0\rangle \in \mathcal{H'}, \: \forall \:t \in \mathbb{Z}^+$.

From this subspace the quantum walk on the tree can be mapped to a quantum walk on a finite line with reflecting boundaries. Define the Hilbert space $\widetilde{\mathcal{H}} = \{ |x,A\rangle : x \in \mathbb{Z}^+, A \in \{L,R\}  \}$ and the subspace $\mathcal{\widetilde{H'}} \subset \mathcal{\widetilde{H}}$ spanned by $\{ |0,L\rangle, |1,R\rangle,|1,L\rangle,\hdots,|n-1,R\rangle,|n-1,L\rangle,|n,R\rangle\}$. Now consider the 1-1 association $\mathcal{H'} \leftrightarrow \mathcal{\widetilde{H'}}$ defined by,
\begin{equation}
|x\rangle_{\text{out}} \leftrightarrow |x, L\rangle, \hspace{0.1in} |x\rangle_{\text{in}} \leftrightarrow |x, R\rangle.
\label{map}
\end{equation}
We also map the operator $U' \leftrightarrow \widetilde{U'}$ so that the action of $U'$ on the states $\{ |x\rangle_{\text{in}},  |x\rangle_{\text{out}}\}$ is equivalent to the action of $\widetilde{U'}$ on the states $\{ |x, R\rangle,|x, L\rangle\}$. The quantum walk on a Cayley tree subject to the above conditions is then seen to be equivalent to a finite one-dimensional quantum walk with a biased coin and asymmetric initial distribution. 

Using the mapping \ref{map}, we can express  \ref{UprimeIn}, \ref{UprimeOut}, \ref{Uprime0}, \ref{Uprime1} in the new basis by defining $\widetilde{U'}:= \widetilde{S}(\mathbbm{1} \otimes \widetilde{C})$, where $\widetilde{C} = \widetilde{C}(x)$ is defined by,
\begin{equation}
(\mathbbm{1} \otimes \widetilde{C})|x,A\rangle = |x\rangle \otimes H(x)|A\rangle.
\end{equation}
\begin{equation*}
\text{ where } H(x) =  \begin{cases}
\:  -\sigma_1 & , \: x=0 \\
\:  h  & , \: 0 < x < n \\
\: \sigma_1& , \: x=n
\end{cases},
\end{equation*}
\begin{equation*}
h = \begin{pmatrix}
      \frac{2\sqrt{d-1}}{d} & \frac{2}{d}-1 \\
     -(\frac{2}{d}-1) & \frac{2\sqrt{d-1}}{d} \\
   \end{pmatrix}\:\text{and }\:
\sigma_1 =  \begin{pmatrix}
   0 & 1\\
   1 & 0\\
   \end{pmatrix}.
\end{equation*}
$\widetilde{S}$ acts on $\mathcal{\widetilde{H'}}$ as follows,
\begin{equation}
\widetilde{S}|x,A\rangle  = \begin{cases}
\: |x+1,R\rangle , & \: A = R \\
\: |x-1, L\rangle , & \: A = L 
\end{cases}.
\end{equation}

It can thus be seen that the quantum walk on the Cayley tree with a central marked vertex starting with equal probability at all vertices is equivalent to a quantum walk on a finite line with a perfectly reflecting boundary at $x=n$ and a $\pi$ phase shift upon reflection at $x=0$. The reflections are acheived by careful choice of the coin operator.

\subsection{Analytical results for $P_s(t)$ and $\omega_s$ for the 2nd generation Cayley Tree}
Consider the 2nd generation $d$-Cayley Tree with a central marked vertex. Let $B$ be a matrix representation of $\widetilde{U'}$ on the complete, orthonormal basis $ \{ |0,L\rangle, |1,R\rangle, |1,L\rangle, |2,R\rangle \} $ of $\widetilde{\mathcal{H'}}$. $\text{Let } |\phi_i\rangle \in \{ |0,L\rangle, |1,R\rangle, |1,L\rangle, |2,R\rangle \}$.
\begin{equation*}
\text{then } \left(B_{ij}\right) = \left( \langle \phi_i | \widetilde{U'} | \phi_j 
\rangle  \right) = \begin{pmatrix}
	0 & \frac{2}{d}-1 & \frac{2\sqrt{d-1}}{d} & 0 \\
	-1 & 0 & 0 & 0 \\
	0 & 0 & 0 & 1 \\
	0 & \frac{2\sqrt{d-1}}{d} & 1-\frac{2}{d} & 0 \\
	\end{pmatrix},
\end{equation*}
where the states are ordered as above. 

We would like to find a closed form expression for $|\Psi(t)\rangle = (\widetilde{U'})^t |\Psi_0\rangle$. In particular we are interested in the probability of finding the walker at the central marked vertex as a function of time. 
\begin{align*}
 |\Psi(t)\rangle = & \: (\widetilde{U'})^t |\Psi_0\rangle \\
 \langle \phi_i | \Psi(t)\rangle = & \:  \sum_j \langle \phi_i | (\widetilde{U'})^t |\phi_j \rangle \langle \phi_j |\Psi_0\rangle,
 \end{align*}
 which can be written in matrix form as,
 \begin{equation}
c_i(t) = (B_{ij})^t c_j(0),
\label{matform}
 \end{equation}
where $c_i(t) :=  \langle \phi_i | \Psi(t)\rangle$ are column vectors. Since $\widetilde{U'}$ is a unitary operator and the basis $|\phi_i\rangle$ is orthonormal, it follows that $B$ is a $4 \times 4$ unitary matrix and therefore has 4 distinct eigenvalues of unit norm corresponding to 4 orthonormal eigenvectors \cite{hassani2006a}. Furthermore, $B$ is unitary similar to the diagonal matrix $D$, which contains the eigenvalues of $B$ as its diagonal elements. Let $P$ be the unitary matrix containing the eigenvectors of $B$ as its columns. From Eq. (\ref{matform}) and the unitarity of P,
 \begin{align*}
{P^{\dagger}}_{ki} c_i(t) = & \: {P^{\dagger}}_{ki}(B_{ij})^t P_{jl}{P^{\dagger}}_{lj} c_j(0),\\
{P^{\dagger}}_{ki} c_i(t) = & \: ({P^{\dagger}}_{ki} B_{ij} P_{jl})^t{P^{\dagger}}_{lj} c_j(0),\\
{P^{\dagger}}_{ki} c_i(t) = & \: (D_{kl})^t{P^{\dagger}}_{lj} c_j(0).
 \end{align*}
 Now writing ${P^{\dagger}}_{ki} c_i(t) =: v_k(t)$ we have,
 \begin{equation}
 v_k(t) = (D_{kl})^t v_l(0).
 \end{equation}
But $D = \text{diag}(\lambda_k)$ where $\lambda_k$ are the eigenvalues of $B$. So $D_{kl}= 0 \text{ for } k \ne l \: \Rightarrow \: v_k(t) = (D_{kk})^t v_k(0)$ or equivalently,
\begin{equation}
v_k(t) = {\lambda_k}^t v_k(0).
\end{equation}
The eigenvalues of B are, 
\begin{equation}
\lambda_k = \pm e^{\pm \Omega/2}, \text{ where } \Omega = \arctan\left(\frac{2\sqrt{d-1}}{d-2}\right).
\label{eigenCT2}
\end{equation}
We can now solve for the time evolution of $v_k(t)$ for $k = 1, \hdots, 4$ and use $P$ to find $c_k(t) :=  \langle \phi_k | \Psi(t)\rangle$, from which we obtain $\langle \mathbbm{1}|\Psi(t)\rangle$.  The eigenvalues are found by diagonalising $B$, \emph{i.e.}
\begin{equation}
D= P^{\dagger}BP = \text{diag}(-e^{-i\Omega/2},-e^{i\Omega/2},e^{i\Omega/2},e^{-i\Omega/2}),
\end{equation}
where the unitary change of basis matrix $P$ contains the orthonormal eigenvectors of $B$ as its columns. 
\begin{equation}
P = \frac{1}{2}\begin{pmatrix}
	i & -i & -i & i \\
	ie^{i\Omega/2} & -ie^{-i\Omega/2} & ie^{-i\Omega/2} & -ie^{i\Omega/2} \\
	-e^{i\Omega/2} & -e^{-i\Omega/2} & e^{-i\Omega/2} & e^{i\Omega/2} \\
	1 & 1 & 1 & 1 \\
	\end{pmatrix}.
\end{equation}
The probability amplitude at the central (marked) vertex is thus
\begin{align*}
\langle \mathbbm{1}|\Psi(t)\rangle \equiv &\: \langle 0, L | \Psi(t)\rangle = c_1(t) = \sum_i P_{1i} v_i(t)\\
=  &\: \frac{1}{2} \bigg[ i(-e^{-i\Omega/2})^t v_1(0) -i(-e^{i\Omega/2})^t v_2(0) \\ 
& - \: ie^{i\Omega t/2}v_3(0) + i e^{-i\Omega t /2}v_4(0)\bigg],
\end{align*}
where $v_i(0)$ represent the initial state in the eigenvector basis and are found using Eq.~\ref{initial} and the unitary transformation $P^\dagger$. Taking the absolute value squared,
\begin{align*}
|\langle \mathbbm{1}|\Psi(t)\rangle|^2 = &\frac{1}{4}\big\{|\mathbf{v}|^2 + 2 \,\mathsf{Re}(v_1\bar{v}_4+\bar{v}_2v_3)\cos(\pi t) \\
& - 2\,\mathsf{Re}(v_1\bar{v}_2+\bar{v}_3v_4)\cos(\Omega t) \\
& - 2\,\mathsf{Im}(\bar{v}_1v_2+v_3\bar{v}_4)\sin(\Omega t) \\
& - 2\,\mathsf{Re}(v_1\bar{v}_3+\bar{v}_2v_4)\cos[(\pi-\Omega)t] \\
& + 2\,\mathsf{Im}(v_1\bar{v}_3+\bar{v}_2v_4)\sin[(\pi-\Omega)t] \big\}.
\end{align*}
We now see that the lowest frequency in the success probability, $\omega_s=\Omega$. The initial condition (Eq.~\ref{initial}) can be converted to the eigenvector basis and the expression (Eq.~\ref{eigenCT2}) for $\Omega(d)$ can then be used to solve for $v_i$. Upon simplification, we obtain Eq.~\ref{PsCT2} (Sec.~\ref{ResAnalytical}), valid for $3 \le d<\infty$.

\subsection{Analytical results for $P_s(t)$ and $\omega_s$ for the 3rd generation Cayley Tree}
We perform the same analysis for the case $n=3$ where this time $\widetilde{U'}$ induces a $6 \times 6$ matrix $B$. We find that the eigenvalues of $B$ are,
\begin{equation}
\lambda_k = \pm i, \pm e^{\pm \Omega/2}, \text{ where } \Omega = \arctan\left(\frac{2\sqrt{d^2-1}}{d^2-2}\right).
\end{equation}
As in the previous case, the eigenvalues of $B$ uniquely determine the frequencies present in $P_s(t)$. Since $i=e^{i\pi/2}$ and $\Omega < \pi/2 \text{ for } 3 \le d < \infty$. It follows that $\omega_s=\Omega$.

\end{document}